# Noise Statistics in Optically Pre-Amplified DPSK Receivers with Optical Mach-Zehnder Interferometer Demodulation

**Xiupu Zhang, and Zhenqian Qu**
*Department of Electrical and Computer Engineering, Concordia University,
Montreal, Quebec, H3G 1M8 CANADA
Tel: 514 848 2424 ext.4107, Fax: 514 848 2802,*
E-mail: xzhang@ece.concordia.ca

**Guodong Zhang**
*AT&T, 200 Laurel Avenue, Middletown, NJ 07748, USA*

**Abstract:** This paper presents for the first time a comprehensive study of noise statistics by use of probability density function (pdf) for DPSK receivers with optical Mach-Zehnder interferometer (MZI) demodulation, considering the impact of signal-amplified spontaneous emission (ASE) beat noise, ASE-ASE beat noise, and phase noise. We further evaluate the error performance of DPSK receivers by using the pdf's and found that balanced detection is less sensitive to phase noise impact than the single-port detection.
© *Optical Society of America*

**References and links:**

1. J. Sinsky A. Adamiccki, A. Gnauck, C. Burrus, J. Leuthold, O. Wohlgemuth, A. Umbach., "A 40Gb/s integrated balanced optical front end and RZ-DPSK performance", *IEEE Photon. Technol. Lett., vol.15, pp.1135-1137, 2003*.
2. A. Gnauck, S. Chandrasekhar, J. Lethold, L. Stulz, "Demonstration of 42.7 Gb/s DPSK receiver with 45 photons/bit sensitivity", *ibid, vol.15, pp.99-101, 2003*.
3. T. Mizuochi, K. Ishida, T. Kobayashi, J. Abe, K. Kinjo, K. Motoshima, K. Kasahara, "A comparative study of DPSK and OOK WDM transmission over transoceanic distances and their performance degradations due to nonlinear phase noise", *J. Lightwave Technol., Vol.21, pp. 1933 - 1942, 2003*.
4. S. Ferber, R. Ludwig, C. Boerner, A. Wietfeld, B. Schmauss, J. Berger, C. Schubert, G. Unterboersch, H. Weber, "Comparison of DPSK and OOK modulation format in 160 Gb/s transmission system", *Electron. Lett., vol.39, pp.1458-1459, 2003*.
5. X. Liu, Y. Kao, M. Movassaghi, R. Giles, "Tolerance to in-band coherent crosstalk of differential phase-shift-keyed signal with balanced detection and FEC", *IEEE Photon. Technol. Lett., vol.16, pp.1209-1211, 2004*.
6. G. Lu, K. Chan, L. Chen, C. Chan, "Robustness of DPSK-WDM systems against nonlinear polarization fluctuation", *ibid, vol.16, pp.927-929, 2004*.
7. P. Winzer, S. Chandrasekhar, H. Kim, "Impact of filtering on RZ-DPSK reception", *ibid vol.15, pp. 840 -842, 2003*.
8. T. Tsuritani, K. Ishida, K. Shimomura, I. Morita, T. Tokura, H. Taga, T. Mizuochi, N. Edagawa, S. Akiba, "70-GHz-spaced 40x42.7 GB/s transpacific transmission over 9400 km using prefilted CSRZ-DPSK signals, all Raman repeaters, and symmetrically dispersion-managed fiber spans", *J. Lightwave Technol., vol.22, pp.215-224, 2004*.
9. J. Wang; J. Kahn, *"*Impact of chromatic and polarization-mode dispersions on DPSK systems using interferometric demodulation and direct detection", *ibid vol.22, pp. 362 – 371 2004*
10. U. Koc; X. Wei, "Combined effect of polarization-mode dispersion and chromatic dispersion on strongly filtered pi/2-DPSK and conventional DPSK", *IEEE Photon. Technol. Lett., vol.16, pp. 1588 – 1590 2004*
11. N. Spellmeyer, J. Gottschalk, D. Caplan, M. Stevens, "High-sensitivity 40-gb/s RZ-DPSK with forward error correction*", ibid, vol.16, pp. 1579 – 1581 2004*
12. X. Wei, A. Gnauck, D. Gill, X. Liu, U. Koc, S. Chandrasekhar, G. Raybon, J. Leuthold, "Optical pi/2-DPSK and its tolerance to filtering and polarization-mode dispersion", *ibid vol.15, pp. 1639 – 1641, 2003*
13. K. Ho**,** "Error probability of DPSK signals with cross-phase modulation induced nonlinear phase noise", *IEEE J. Selected Topics in Quantum Electronics, vol.10, pp. 421 – 427 2004*
14. H. Kim, P. Winzer, "Robustness to laser frequency offset in direct-detection DPSK and DQPSK systems", *J. Lightwave Technol., vol.21, pp.1887 – 1891 2003*
15. J. Wang, J. Kahn, **"**Accurate bit-error-ratio computation in nonlinear CRZ-OOK and CRZ-DPSK systems", *IEEE Photon. Technol. Lett., vol.16, pp. 2165 – 2167 2004*
16. H. Kim, "Cross-phase-modulation-induced nonlinear phase noise in WDM direct-detection DPSK systems", *J. Lightwave Technol., vol.21, pp.1770-1774 2003*

17. C. Xu, X. Liu; L. Mollenauer, X. Wei, "Comparison of return-to-zero differential phase-shift keying and ON-OFF keying in long-haul dispersion managed transmission", *IEEE Photon. Technol. Lett., vol.15, pp.617-619 2003*
18. C. Xu, X. Liu, X. Wei, "Differential phase-shift keying for high spectral efficiency optical transmissions", *IEEE J. Selected Topics in Quantum Electronics, vol.10, pp. 281 – 293 2004*
19. B. Zhu, L. Nelson, S. Stulz, A. Gnauck, C. Doerr, J. Leuthold, L. Gruner-Nielsen, M. Pedersen, J. Kim, R. Lingle, "High spectral density long-haul 40-Gb/s transmission using CSRZ-DPSK format", *J. Lightwave Technol. vol.22, pp. 208 – 214, 2004*
20. C. Rasmussen, T. Fjelde, J. Bennike, F. Liu; S. Dey, B. Mikkelsen, P. Mamyshev, P. Serbe, P. Wagt, Y. Akasaka, D. Harris, D. Gapontsev, V. Ivshin, P. Reeves-Hall, "DWDM 40G transmission over trans-pacific distance (10 000 km) using CSRZ-DPSK, enhanced FEC, and all-Raman-amplified 100-km ultrawave fiber spans", *ibid vol.22, pp. 203 – 207, 2004*
21. X. Wei; X. Liu; C. Xu, "Numerical simulation of the SPM penalty in a 10-Gb/s RZ-DPSK system", *IEEE Photon. Technol. Lett., vol.15, pp. 1636 – 1638, 2003.*
22. G. Bosco, P. Poggiolini, "On the Q factor inaccuracy in the performance analysis of optical direct-detection DPSK systems", *ibid, vol.16, pp. 665 – 667, 2004*
23. A. Agarwal, S. Banerjee, D. Grosz, A. Kung, D. Maywar, T. Wood, "Ultralong-haul transmission of 40-Gb/s RZ-DPSK in a 10/40 G hybrid system over 2500 km of NZ-DSF", *ibid vol.15, pp. 1779 – 1781 2003*
24. P. Smith, M. Shafi, C. Kaiser, "Optical heterodyne binary DPSK systems: a review of analysis and performance", *IEEE J. Selected Areas in Communications, vol.3, pp.525-533, 1995.*
25. O. Tonguz, R. Wagner, "Equivalence between pre-amplified direct detection and heterodyne receivers", *IEEE Photon. Technol. Lett., vol.3, pp.835-837, 1991.*
26. G. Jacobsen, I. Garrett, "Theory for optical heterodyne DPSK receivers with post-detection filtering", *J. Lightwave Technol., vol.5 pp.478-484, 1987.*
27. S. Betti, G. Marchis, E. Iannone, "Coherent optical communication systems," John Wiley & Sons, 1995.
28. X. Zhang, G. Zhang, C. Xie, L. Wang, "Noise statistics in optically pre-amplified DPSK receivers with Mach-Zehnder interferometer demodulation", *Opt. Lett., vol.29, pp.337-339, 2004.*
29. Y. Yadin, M. Shtaif, M. Orenstein, "Nonlinear phase noise in phase-modulated WDM fiber optic communications", *IEEE Photon. Technol. Lett., vol.16, pp.1307-1309, 2004.*
30. K. Ho, "Performance degradation of phase-modulated systems due to nonlinear phase noise", *ibid, vol.15, pp.1213-1215, 2003.*
31. A. Mecozzi, "Probability density functions of the nonlinear phase noise", *Opt. Lett., vol.29, pp.673-675, 2004.* (if the nonlinear phase noise $\Phi(t)$ has the Gaussian distribution the phase difference $\Delta\Phi = \Phi(t) - \Phi(t-T_b)$ must be the Gaussian.)
32. K. Ho, "Probability density function of nonlinear phase noise", *J. Opt. Soc. Am. B, vol.20 pp.1875-1879 2003.*
33. K. Ho, Asymptotic probability density of nonlinear phase noise", *Opt. Lett., vol.28 pp.1350-1352, 2003.*
34. K. Ho, "Impact of nonlinear phase noise to DPSK signals: a comparison of different models", *IEEE Photon. Technol. Lett., vol.16, pp.1403-1405, 2004.*
35. P. Winzer, H. Kim, "Degradations in balanced DPSK receivers", *ibid, vol.15 pp.1282-1284, 2003.*
36. A. Gnauck and P. Winzer, "40Gb/s RZ-differential phase shift keyed transmission", *OFC'2004, Paper TuF5, 2004.*
37. P. Humblet, M. Azizoglu, "On the bit error rate of lightwave systems with optical amplifiers", *J. Lightwave Technol., vol.9, pp.1576-1582, 1991.*
38. G. Agrawal, *Fiber-Optic Communication Systems,* 3rd ed., John Wiley & Sons, Inc., New York, 2002, pp.261.
39. D. Marcuse, "Derivation of analytical expressions for the bit-error probability in lightwave systems with optical amplifiers", *J. Lightwave Technol., vol.8, pp. 1816-1823, 1990.*
40. D. Marcuse, "Calculation of bit-error probability for a lightwave system with optical amplifiers and post-detection Gaussian noise", *J. Lightwave Technol., vol.9, pp.505-513, 1991.*
41. S. Chinn, D. Boroson, J. Livas, "Sensitivity of optically pre-amplified DPSK receivers with Fabry-Perot filters", J. *Lightwave Technol., vol.14 pp.370-375, 1996.*
42. A. Papoulis, S. Pillai, *Probability, Random variables and Stochastic processes*, McGraw-Hill, 2002, pp.181-186.
43. H. Kim, A. Gnauck, "Experimental investigation of the performance limitation of DPSK systems due to nonlinear phase noise", *IEEE Photon. Technol. Lett., vol.15, 320-322, 2003.*

## 1. Introduction

In order to utilize the transmission bandwidth more efficiently and to increase the aggregated capacity, the bit rates of 40 Gb/s and beyond per channel with the channel space of ≤ 50 GHz are attractive for applications into long-haul and ultra long-haul wavelength division multiplexing (WDM) transmission systems. However, for such systems with the traditional intensity modulation/direct detection (IM/DD), it has been found that several impairments, such as cross-phase modulation, inter- and intra-channel crosstalk and optical filtering etc., become very severe and significantly degrade the transmission performance. In order to suppress these impairments, several novel modulation formats have been proposed and investigated, such as the carrier-suppressed non return-to-zero, carrier-suppressed return-to-zero, duobinary, and differential phase-shift keying (DPSK), etc. Among these modulation formats, DPSK has attracted great attention and been investigated extensively [1-23]. This is because DPSK has several advantages over IM/DD: DPSK can provide additional 3-dB receiver sensitivity or Q factor [1-4], and is more tolerant of signal power fluctuation, in-band crosstalking [5], nonlinear polarization fluctuation [6] and optical filtering [7].

DPSK receivers with coherent detection (homodyne and heterodyne) and electrical demodulation (referred to as conventional DPSK receivers) were investigated extensively [24-27]. However, DPSK with optical Mach-Zehnder interferometer (MZI) demodulated receivers (referred to as DPSK/MZI receivers) is different from the conventional DPSK receivers in the physical processing of signal and noise. In the conventional DPSK receivers, the electrical demodulation is after optical detection, and DPSK/MZI receivers have the demodulation before the optical detection. This implies that the theory in [24-27] could not be directly applied for the DPSK/MZI receivers with balanced detection.

In order to analyze the performance of DPSK systems, it is essential to know the probability density function (pdf) of noise statistics in DPSK/MZI receivers. However, the noise statistics in optically pre-amplified DPSK/MZI receivers have not been fully investigated and understood theoretically. So far, either noise statistics of the phase noise or the amplified spontaneous emission noise (ASE) is investigated separately [7,13, 28-35]. A comprehensive analysis is needed to explore the noise statistics of DPSK/MZI receivers.

## 2. Noise Statistics in DPSK/MZI Receivers

Since the MZI before the optical detector converts the phase modulated optical signal into intensity-modulated signal, the optical detection and electrical processing in DPSK receivers are the same as those in IM/DD receivers. Then, similar to optically pre-amplified IM/DD receivers [39-40] the shot noise, thermal noise, signal-ASE beat noise, and ASE-ASE beat noise (besides the phase noise) are involved in optically pre-amplified DPSK receivers. The phase noise may come from laser phase noise, quadrature component of ASE noise and nonlinear phase noise. In this work, we intend to analyze the pdf's of noise statistics in optically pre-amplified DPSK/MZI receivers with single port and balanced detection. Due to the fact that the shot noise and thermal noise are negligibly small compared to the signal-ASE beat noise and ASE-ASE beat noise in optically pre-amplified receivers, these two noise contributions are not considered here.

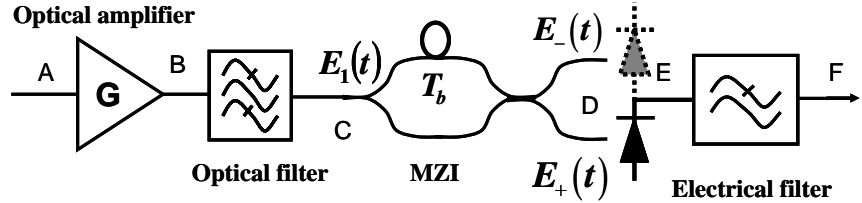

Fig.1: Basic block structure of optically pre-amplified DPSK/MZI receivers.

Fig. 1 depicts the schematic structure of DPSK/MZI receivers. Compared to IM/DD receivers, the only difference is that a MZI demodulator is inserted before the optical detector, and the MZI is a special optical filter. DPSK/MZI receivers with constructive port, destructive port, and balanced detection are corresponding to the structure shown in Fig.1 with the black, gray, and both photodiodes together, respectively.

Assuming that two power splitters of the MZI are ideal and have the exact 3-dB splitting power ratio, we can obtain a relation of the electric fields between input and output ports of the MZI as $E_+(t) = -j/2 \times [E_1(t-T_b) + E_1(t)]$, and $E_-(t) = [E_1(t-T_b) - E_1(t)]/2$, where $T_b$ is the bit period, $E_+(t)$ and $E_-(t)$ stand for electric fields of constructive and destructive ports of MZI, respectively, and $E_1(t)$ denotes the input electric field of the MZI. We express the electric field at the point C in Fig.1 by $E_1(t) = E_s(t)\exp[j\theta(t) + j\Phi(t)] + n(t)$ where $E_s(t)$ represents the amplitude and $\theta(t)$ the phase modulation of DPSK signal, $\Phi(t)$ is the phase noise, and $n(t)$ stands for band-limited ASE noise. The currents from constructive and destructive ports (the point E in Fig.1) are given by

$$I_+(t) = R|E_+(t)|^2 = R|E_{S+}(t) + n_+(t)|^2 = I_S/2 \times [1 + \cos(\Delta\theta + \Delta\Phi)] + Rn_{SA}^+ + Rn_{AA}^+ \quad (1a),$$

$$I_-(t) = R|E_-(t)|^2 = R|E_{S-}(t) + n_-(t)|^2 = I_S/2 \times [1 - \cos(\Delta\theta + \Delta\Phi)] + Rn_{SA}^- + Rn_{AA}^- \quad (1b)$$

where $n_+(t)$ and $n_-(t)$ represent the output ASE noise at constructive and destructive ports, respectively, $\Delta\theta = \theta(t - T_b) - \theta(t)$ and $\Delta\Phi = \Phi(t - T_b) - \Phi(t)$. In (1), the condition of $E_s(t - T_b) = E_s(t) = E_s$ and $I_s = R|E_s|^2 = RP_s$ were applied ($P_s$ is the average optical power of the signal, and $n_{SA}^+(t)$ ($n_{AA}^+(t)$) and $n_{SA}^-(t)$ ($n_{AA}^-(t)$) stand for the signal-ASE beat noise (ASE-ASE beat noise) occurring at the constructive and destructive port, respectively).

The autocorrelation function is defined as $R(\tau) = \langle n(t)n^*(t-\tau) \rangle$. Thus the noise correlation function is given by $R(\tau) = N_{ASE}B_o \sin(\pi B_o \tau)/(\pi B_o \tau)$ for the ideal optical filter with the rectangular spectral shape, and $R(\tau) = N_{ASE}B_o \exp[-\pi B_o \tau]$ for the Fabry-Perot optical filter [41]. For DPSK, the time delay of $\tau = T_b$ is used. The correlation factors of $|\sin(\pi B_o T_b)/(\pi B_o T_b)|$ and $\exp[-\pi B_o T_b]$ are less than 13% and 0.2% for $B_o T_b \geq 2$. Then, the correlation of the ASE noise $n(t)$ and $n(t - T_b)$ can be ignored without loss of accuracy. For the ideal DPSK, i.e., without any phase error, we can obtain from (1) that if $\Delta\theta = 0$ for bit "1", $I_+(t) \sim I_s$, and $I_-(t) \sim 0$; and if $\Delta\theta = \pi$ for bit "0", $I_+(t) \sim 0$ and $I_-(t) \sim I_s$. In other words, if bit "1" is received, the optical signal will completely present at the constructive port and only ASE noise will present at the destructive port; if bit "0" is received, the optical signal will completely appear at the destructive port and only ASE noise will appear at the constructive port. However, as the phase noise is induced, the above facts may not be held.

Finally for DPSK/MZI receivers with constructive port detection, the currents for bits "1" and "0" are given from (1a) by

$$I_1(t) = I_+(t) = I_s/2 \times [1 + \cos(\Delta\Phi)] + Rn_{SA}^+(t) + Rn_{AA}^+(t) \quad (2a),$$

$$I_0(t) = I_+(t) = I_s/2 \times [1 - \cos(\Delta\Phi)] + Rn_{SA}^+(t) + Rn_{AA}^+(t) \quad (2b)$$

It can be seen that the signal-ASE beat noise occurs in bit "0" if phase error occurs. But the signal-ASE beat noise is weak compared to that in bit "1". For the DPSK/MZI receivers with destructive port detection, the currents for bits "1" and "0" are given from (1b) by

$$I_1(t) = -I_-(t) = -I_s/2 \times [1 - \cos(\Delta\Phi)] - Rn_{SA}^-(t) - Rn_{AA}^-(t) \quad (3a),$$

$$I_0(t) = -I_-(t) = -I_s/2 \times [1 + \cos(\Delta\Phi)] - Rn_{SA}^-(t) - Rn_{AA}^-(t) \quad (3b)$$

Expression (3) shows that the weak signal-ASE beat noise could occur in bit "1" as well. For DPSK/MZI receivers with balanced detection, the currents for bits "1" and "0" are given from (1) by

$$I_1(t) = I_+(t) - I_-(t) \approx I_s \cos(\Delta\Phi) + Rn_{SA}^+(t) + Rn_{AA}^+(t) - Rn_{AA}^-(t) \quad (4a),$$

$$I_0(t) = I_+(t) - I_-(t) \approx -I_s \cos(\Delta\Phi) - Rn_{SA}^-(t) - Rn_{AA}^-(t) + Rn_{AA}^+(t) \quad (4b)$$

For simplicity, it is assumed, for expressions (4), that the signal-ASE beat noise presents only at the constructive port and does not occurs at the destructive port for bit "1", and vice versa for bit "0". In the following, we separately discuss the noise statistics for three cases of DPSK/MZI receivers: the constructive single port, the deconstructive single port, and balanced detection.

### 2.1 DPSK/MZI receivers with constructive port detection

We follow the same procedures as in IM/DD receivers [39-40]. The conditional pdf for bit "1" is given, based on (2), by

$$f_1(x|\Delta\Phi) = \frac{M}{\overline{I}_{A+}} \left(\frac{x}{\overline{I}_1}\right)^{\frac{M-1}{2}} \exp\left(-M\frac{x+\overline{I}_1}{\overline{I}_{A+}}\right) I_{M-1}\left(\frac{2M\sqrt{x\overline{I}_1}}{\overline{I}_{A+}}\right), x \geq 0 \qquad (5)$$

where $M = B_o/B_e$, $\overline{I}_1 = I_s/2 \times [1+\cos(\Delta\Phi)]$ as the signal current of bit "1", and $\overline{I}_{A+} = 2RN_{ASE}B_+$ as the average current generated by ASE noise at constructive port, and $B_+ = B_o/2 + B_o \sin(\pi B_o T_b)/(2\pi B_o T_b)$ as the equivalent optical noise bandwidth of the constructive port, obtained for the ideal optical filter with the rectangular spectral shape. For bit "0", the pdf is obtained by replacing $\overline{I}_1$ by $\overline{I}_0 = I_s/2 \times [1-\cos(\Delta\Phi)]$ in (5) if $\Delta\Phi \neq 0$.

For $\Delta\Phi = 0$, the pdf for bit "0" is reduced to $f_0(x) = \left(\frac{M}{\overline{I}_{A+}}\right)^M \frac{x^{M-1}}{\Gamma(M)} \exp\left[-M\frac{x}{\overline{I}_{A+}}\right], x \geq 0$.

Then we consider the influence of phase noise. It has been shown that phase difference $\Delta\Phi$ can be well approximated by the Gaussian distribution [29-33], i.e. $f_{\Delta\Phi}(\Delta\Phi) = 1/\sqrt{2\pi\sigma_{\Delta\Phi}^2} \exp\left[-\Delta\Phi^2/2\sigma_{\Delta\Phi}^2\right]$, where $\sigma_{\Delta\Phi}^2$ is the variance of the phase difference of the phase noise. By using the total probability, the pdf of the noise statistics for bits "1" is given by

$$f_1(x) = \int_{-\infty}^{\infty} f_{\Delta\Phi}(\Delta\Phi) f_1(x|\Delta\Phi) d\Delta\Phi \qquad (6).$$

Expression (6) presents the pdf of noise statistics in DPSK/MZI receivers with constructive port detection, the pdf for bit "0" can be derived in a similar way as that for expression (6).

## 2.2 DPSK/MZI receivers with destructive port detection

We follow the same procedures as the above. The conditional pdf based on (1b) for bit "1" is given by

$$f_1(x|\Delta\Phi) = \frac{M}{\overline{I}_{A-}} \left(\frac{x}{\overline{I}_1}\right)^{\frac{M-1}{2}} \exp\left(-M\frac{x+\overline{I}_1}{\overline{I}_{A-}}\right) I_{M-1}\left(\frac{2M\sqrt{x\overline{I}_1}}{\overline{I}_{A-}}\right), x \leq 0 \qquad (7)$$

where $\overline{I}_1 = -1/2 I_s [1-\cos(\Delta\Phi)]$ as the signal current of bit "1", and $\overline{I}_{A-} = 2RN_{ASE}B_-$ as the average current generated by destructive port ASE noise, and $B_- = B_o/2 - B_o \sin(\pi B_o T_b)/(2\pi B_o T_b)$ as the equivalent optical noise bandwidth of destructive port, obtained for the ideal optical filter with the rectangular spectral shape. The pdf for bit "0" can be obtained by replacement of $\overline{I}_1$ with $\overline{I}_0 = -1/2 I_s [1+\cos(\Delta\Phi)]$. Thus, the exact pdf's are obtained by (6) with the help of (7). For the case of $\Delta\Phi = 0$, the pdf for bit "0" is reduced to $f_0(x) = \left(\frac{M}{\overline{I}_{A-}}\right)^M \frac{x^{M-1}}{\Gamma(M)} \exp\left[-M\frac{x}{\overline{I}_{A-}}\right], x \geq 0$.

## 2.3 DPSK/MZI receivers with balanced detection

We have seen in (4) that the currents for bits "1" and "0" contain noise from both constructive and destructive ports. The conditional pdf for the first three terms in (4a) can be given by

$$f_+(x|\Delta\Phi) = \frac{M}{\overline{I}_{A+}} \left(\frac{x}{\overline{I}_1}\right)^{\frac{M-1}{2}} \exp\left(-M\frac{x+\overline{I}_1}{\overline{I}_{A+}}\right) I_{M-1}\left(\frac{2M\sqrt{x\overline{I}_1}}{\overline{I}_{A+}}\right), x \geq 0, \text{ where } \overline{I}_1 = I_s \cos(\Delta\Phi)$$

as the signal current of bit "1", and $\bar{I}_{A+} = 2RN_{ASE}B_+$. The pdf for the last term in (4a), which is from the destructive port, is given by $f_-(y) = \left(\dfrac{M}{\bar{I}_{A-}}\right)^M \dfrac{y^{M-1}}{\Gamma(M)} \exp\left[-M\dfrac{y}{\bar{I}_{A-}}\right], y \geq 0$, where $\bar{I}_{A-} = 2RN_{ASE}B_-$. Assuming the linear ASE noises at the constructive and destructive ports are independent [34], then, the conditional pdf for bit "1" is obtained based on (4a) by [42]

$$f_1(x|\Delta\Phi) = \int_0^\infty f_+(x+y|\Delta\Phi) f_-(y) dy \qquad (8).$$

Consequently, the totally pdf for bit "1" can be obtained by (6) with the use of (8). Similarly the totally pdf for bit "0" can also be obtained using the same procedures based on (4b).

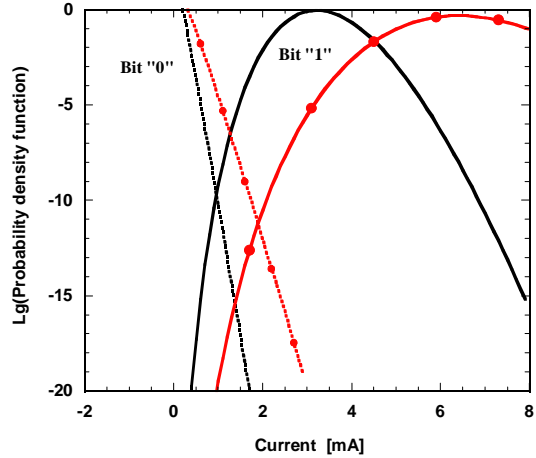

Fig. 2: pdf's of bits "1" and "0" for IM/DD (red) and DPSK/MZI receivers with constructive port detection without phase noise (black). Bit "1": solid, and bit "0": dashed.

## 3. Numerical Calculations of pdf's

For DPSK/MZI receivers with constructive port detection, the detection currents for the case of no phase error are given by $I_1(t) = I_s + Rn^+_{SA}(t) + Rn^+_{AA}(t)$ and $I_0(t) = Rn^+_{AA}(t)$. The currents are similar to those in IM/DD receivers except the average current of bit "1". Fig. 2 depicts the pdf's. The parameters which used for the calculation are: bit rate of 43 Gb/s; optical pre-amplifier gain of 35 dB, noise figure of 6 dB, the average optical signal power of -30 dBm, R=1, $B_o$ =100 GHz, and $B_e$ =33 GHz.

For DPSK/MZI receivers with destructive port detection, the detection currents for the case of no phase error are given by $I_1(t) = -I_s - Rn^-_{SA}(t) - Rn^-_{AA}(t)$ and $I_0(t) = -Rn^-_{AA}(t)$. Similar to Fig. 2, the pdf's are shown in Fig. 3. The small difference between the pdf's for constructive and destructive port detections is the resulting from that the contribution of ASE-ASE beat noise in constructive port detection is different from that in destructive port detection.

For DPSK/MZI receivers with balanced detection, the detection currents for the case of no phase error are given by $I_1(t) = I_s + Rn^+_{SA}(t) + Rn^+_{AA}(t) - Rn^-_{AA}(t)$ and $I_0(t) = -I_s - Rn^-_{SA}(t) - Rn^-_{AA}(t) + Rn^+_{AA}(t)$. In $I_1(t)$, the first three terms are introduced by the constructive port only, and the last term is from the destructive port only. For the ideally balanced detection, the pdf's for bits "1' and "0" are symmetrical. We illustrate the calculated pdf's in Fig. 3. In $I_1(t)$ and $I_0(t)$, the last two terms mean the ASE-ASE beat noise. If these two terms are ignored the pdf's of $I_1(t)$ and $I_0(t)$ become the Gaussian distributed.

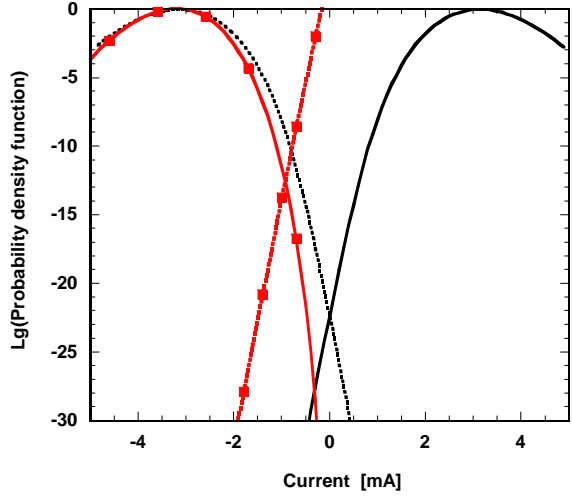

Fig. 3: pdf's of bits "1" and "0" for DPSK/MZI receivers with destructive port (red) and balanced detection (black) without phase noise. The parameters used are the same as in Fig.2. Bit "1": solid, and bit "0": dashed.

Now we analyze the impact of the phase noise on pdf's. For DPSK/MZI receivers with constructive port detection, we illustrate the pdf's in Fig. 4 for $\sigma_{\Delta\Phi} = 0$, 0.1, 0.2, 0.3, and 0.5 radians. It is obvious that the pdf for bit "0" is very sensitive to phase noise; and in contrast the pdf for bit "1" is much less sensitive to phase noise (two pdf's for $\sigma_{\Delta\Phi} = 0$, and 0.1 are not distinguishable for bit "1" in Fig. 4). Moreover, as $\sigma_{\Delta\Phi}$ increases, the cross point of bits "1" and "0" pdf's is increased due to the phase noise's influence on the pdf of bit "0". Furthermore, the optimal decision threshold is very sensitive to the phase noise. Since DPSK/MZI receivers with constructive and destructive port detection are almost the same except a slight difference in the ASE-ASE beat noise, the behavior of the phase noise influence on the pdf's for the destructive port detection is expected to be very similar to that for constructive port detection.

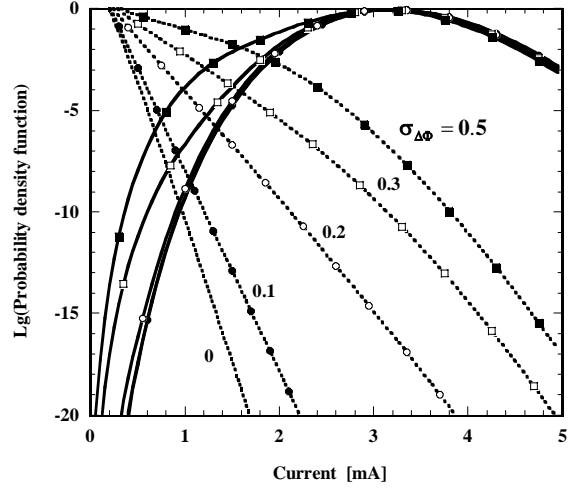

Fig. 4: pdf's of bits "1" and "0" for DPSK/MZI receivers with constructive port detection for $\sigma_{\Delta\Phi} = 0$, 0.1, 0.2, 0.3 and 0.5 radians. Bit "1" (solid) and bit "0" (dashed). The other parameters used are the same as in Fig. 2.

For DPSK/MZI receivers with balanced detection, the pdf's are presented in Fig. 5. The shape of the pdf's in Fig. 5 is in good agreement with the measured one [43]. Due to the symmetry of the pdf's of bits "1" and "0", the impact of phase noise on the pdf's are the same for both the bit "1" and "0". Consequently, we can expect, to some extent, that the influence of the phase noise on DPSK/MZI receivers with balanced detection is less than that on DPSK/MZI receivers with single port detection. For example, the optimum decision threshold is almost independent on the phase noise as seen in Fig. 5. If ignoring the ASE noise in the DPSK/MZI receivers, the detection currents become $2I_1(t) - I_s = I_s \cos(\Delta\Phi)$, and $2I_0(t) - I_s = -I_s \cos(\Delta\Phi)$ for the single port detection, obtained from (2); and $I_1(t) = I_s \cos(\Delta\Phi)$, and $I_0(t) = -I_s \cos(\Delta\Phi)$ for balanced detection, obtained from (4). These currents show that the maximum phase noise impact is the same in single port and balanced detection.

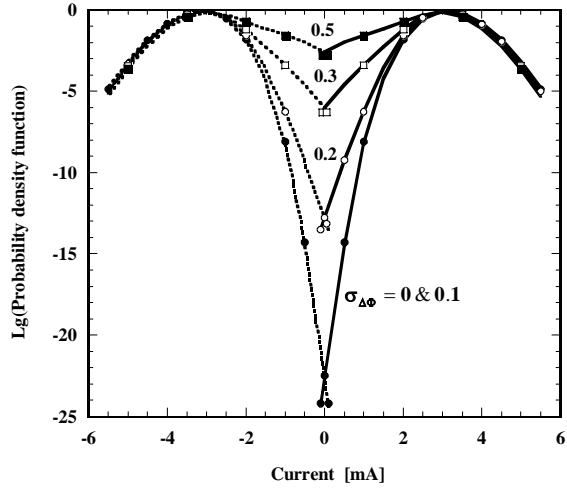

Fig. 5: pdf's of bits "1" and "0" for DPSK/MZI receivers with balanced detection. The other parameters used are the same as in Fig. 4.

## 4. Comparisons of DPSK/MZI Receivers

In this section, we use the cumulative probability (CP) to evaluate the performance of DPSK/MZI receivers with single port and balanced detection. The CP is defined as $CP = \frac{1}{2}\left[\int_{-\infty}^{I_{th}} f_1(x)dx + \int_{I_{th}}^{\infty} f_0(x)dx\right]$ where $f_1(x)$ and $f_0(x)$ are the total pdf's of bits '1" and "0", and $I_{th}$ is the optimal decision threshold of the detection currents. The CP is considered as the bit error ratio (BER) in IM/DD [37-40] and DPSK/MZI receivers [7, 9, 13, 15, 29-30, 34-37].

Fig. 6 shows the CP for IM/DD receivers (red), DPSK/MZI receivers with constructive port detection (blue) and destructive port detection (black) with the use of pdf's. The results in Fig. 6 are in good agreement with those in Ref. [35] where the performance of DPSK/MZI receivers with constructive and destructive port detection is not identical. Again, the performance difference is due to the ASE-ASE beat noise in the two receivers.

The performance comparison of DPSK/MZI receivers with balanced detection to IM/DD receivers is illustrated in Fig. 7. It is observed that DPSK/MZI receivers with balanced detection have ~3 dB improvement of optical receiver sensitivity compared to IM/DD receivers for the same CP. This agrees well with the results in [7, 35, 37] even though the procedures of calculations are quite different from this work, which verifies that our calculated pdf's for balanced detection is correct and accurate. Our further calculations have shown that the receiver sensitivity improvement for balanced detection over single port detector decreased as the factor of $B_o/B_e$ is increased, and then completely vanished at a high ration of $B_o/B_e$. This finding, again, agrees well with the observations in Refs. [7, 37].

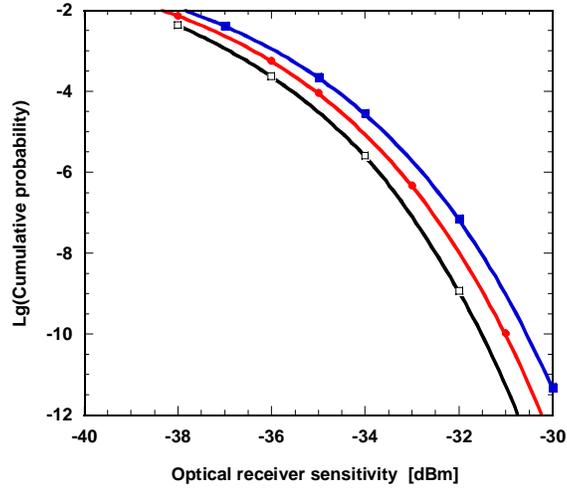

Fig. 6: CP as a function of optical receiver sensitivity for IM/DD receivers (red) and DPSK/MZI receivers with constructive (blue) and destructive (black) port detection without phase noise. The other parameters used are the same as in Fig. 2.

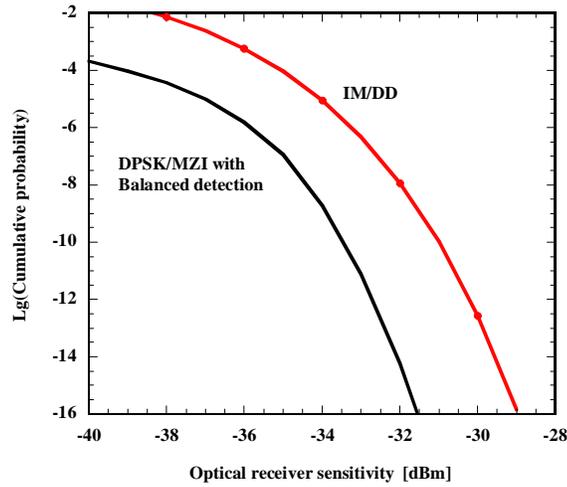

Fig. 7: CP as a function of optical receiver sensitivity for IM/DD receivers (red) and DPSK/MZI receivers with balanced detection (black) without phase noise. The other parameters used are the same as in Fig. 2.

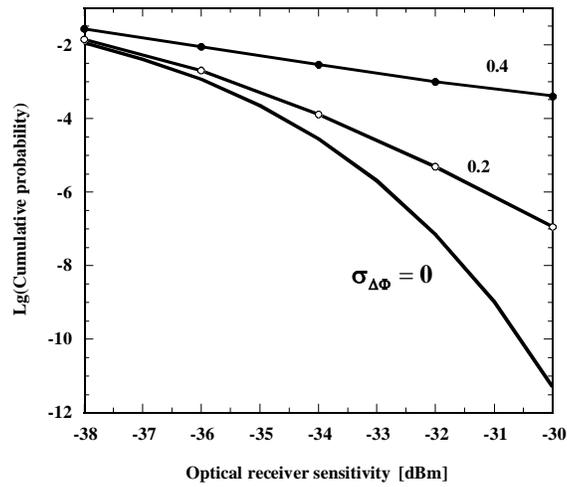

Fig. 8: CP as a function of receiving power for DPSK/MZI receivers with constructive port detection for $\sigma_{\Delta\Phi} = 0$, 0.2 and 0.4 radians.

Next we investigate the impact of phase noise on CP in DPSK/MZI receivers. Fig. 8 shows CP with $\sigma_{\Delta\Phi} = 0$, 0.2 and 0.4 radians for constructive port detection. As seen in the figure, the CP increases very rapidly as the phase noise increases. Moreover, the error floor of CP is presented when $\sigma_{\Delta\Phi} = 0.4$ radians.

Fig. 9 shows the impact of the phase noise on DPSK/MZI receivers with balanced detection. The CP also increases rapidly as the phase noise increases, but is much smaller than the CP for constructive port detection at a given phase noise. In this case the floor of CP occurs when $\sigma_{\Delta\Phi} = 0.4$ as well. Then, there is about the same upper limit of phase noise impact for both single-port and balanced detections, which agrees with the results in Ref. [43]. However, the receiver with balanced detection is less sensitive to the phase noise than that with single port detection if below the upper limit.

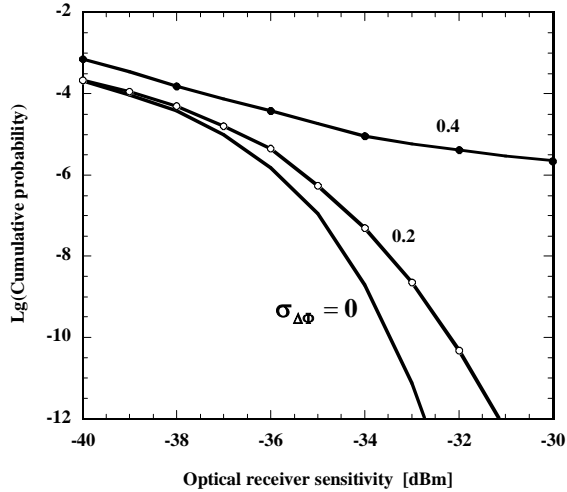

Fig. 9: CP as a function of receiving power for DPSK/MZI receivers with balanced detection.

## 5. Conclusions

We have presented a comprehensive analysis of noise statistics by use of pdf's for DPSK/MZI receivers with constructive, destructive port and balanced detection, considering the impact of signal-ASE beat noise, ASE-ASE beat noise, and phase noise.

We have found that, if the standard deviation of differential phase noise is 0.4 radians and beyond, the similar noise floor occurs for both single port and balanced detection. On the other hand, it is also shown that DPSK/MZI receivers with balanced detection are less sensitive to phase noise than single port detection if below 0.4 radians.

Reviewer 1 comments:

This paper adds relatively little to the existing literature on the topic of the nonlinear phase noise in DPSK-format lightwave systems. It also has a number of incorrect statements. It is recommended that this paper should not be accepted in Optics Express.

Reviewer 2 comments:

The paper should be rejected because it does not provide any new understanding to the problem about the performance of DPSK signal with nonlinear phase noise when interferometer is used at the receiver. It seems that the authors do not fully understand the mathematics behind the problem.

First of all, in the third pages, the last two paragraphs of Section 1 are totally wrong. If the authors are not able to appreciate and understand the results of other previous literatures, I really don't know that they are able to handle the mathematics of the problems correctly.

1. The authors said that "the theory in [24-27] could not be directly applied for the DPSK/MZI receivers". Please note that the title of [25] is "Equivalence between pre-amplified direct detection and heterodyne receivers". For two mathematically equivalence problems, the theory is [24-27] certainly can directly applied for the DPSK/MZI receivers. The authors must first prove that the paper of [25] is wrong. Unfortunately, there is no effort by the author to clarify this confusion.

2. The authors also said that "So far, either noise statistics of the phase noise or the amplified spontaneous emission noise (ASE) is investigated separately [7,13, 28-35]". Some of those papers calculate the error probability, may be using simplification or approximation. In order to calculate the error probability, both ASE and phase noise should be taken into account.

With all papers of [24-27] and [7, 13, 28-35], I don't believe that the current paper contributes anything new to the topic. About the phase noise, some of those previous papers consider phase noise to be non-Gaussian distributed, as confirmed by measurement. The current paper compares the theory with measurement but roughly looking at the shape of the probability distribution. The current paper maintains that the phase noise is Gaussian distributed that is the same as laser phase noise. If phase noise is Gaussian distributed, there is on reason we need this new paper with all the analysis in [24, 26, 27] for laser phase noise and the argument in [25] of the equivalence. If phase noise is non-Gaussian distributed, this paper is wrong. In both cases, this paper should be rejected.

Currently, there is still argument on whether the phase noise is Gaussian or non-Gaussian distributed when [31] and [33] are compared. However, the numerical results in [34] seem to indicate that the assumption of non-Gaussian distribution gives worse error probability and larger penalty. While it is arguable whether the difference of only up to fraction of a dB in [34] is significant, just completely ignores the issue in the current paper is not appropriate.